\documentclass[twocolumn,aps,superscriptaddress,preprintnumbers]{revtex4}

\usepackage{inputenc}
\usepackage{supertabular}
\usepackage{graphicx}
\usepackage{dcolumn}
\usepackage{bm}
\usepackage{color}
\usepackage{textcomp}
\usepackage{amsmath}
\usepackage{amssymb}
\usepackage{braket}
\usepackage{amsmath,times,mathtools}
\usepackage{graphicx}
\usepackage{physics}
\usepackage{hyperref}
\usepackage{setspace}
\usepackage{tabularx}
\usepackage{xpatch}%
\usepackage{multirow}
\usepackage{setspace}
\usepackage{amsmath,times,mathtools}
\usepackage{graphicx}
\usepackage{physics}
\usepackage{hyperref}
\usepackage{setspace}
\usepackage{tabularx}
\usepackage{adjustbox}
\usepackage{xcolor}
\usepackage[sectionbib]{bibunits}
\usepackage[final]{pdfpages}

\usepackage{hyperref}
\hypersetup{
    colorlinks=true,       
    linkcolor=blue,        
    citecolor=blue,        
    filecolor=blue,        
    urlcolor=blue,         
    runcolor=blue
}

\begin{document}
\title{Giant Generation of Polarization-Entangled Photons in Metal Organic Framework Waveguides}

\author{Simón Paiva}
\affiliation{Department of Physics, Universidad de Santiago de Chile, Av. Victor Jara 3493, Santiago,Chile}

\author{Ruben A. Fritz }
\affiliation{Department of Physics, Universidad de Santiago de Chile, Av. Victor Jara 3493, Santiago,Chile}

\author{Sanoj Raj}
\affiliation{Department of Chemical and Biomolecular Engineering, University of Notre Dame, 
IN 46556, USA}

\author{Yamil J. Col\'on}
\affiliation{Department of Chemical and Biomolecular Engineering, University of Notre Dame, 
IN 46556, USA}

\author{Felipe Herrera}
\affiliation{Department of Physics, Universidad de Santiago de Chile, Av. Victor Jara 3493, Santiago,Chile}
\affiliation{Millennium Institute for Research in Optics, Concepción, Chile}
\email{felipe.herrera.u@usach.cl}

\begin{abstract} 
Parametric nonlinear optical processes are instrumental in optical quantum technology for generating entangled light. However, the range of materials conventionally used for producing entangled photons is limited. Metal-organic frameworks (MOFs) have emerged as a novel class of optical materials with customizable nonlinear properties and proven chemical and optical stability. The large number of combinations of metal atoms and organic ligand from which bulk MOF crystals are known to form, facilitates the search of promising candidates for nonlinear optics. To accelerate the discovery of next-generation quantum light sources, we employ a multi-scale modeling approach to study phase-matching conditions for collinear degenerate type-II spontaneous parametric down conversion (SPDC) with MOF-based one dimensional waveguides. Using periodic-DFT calculations to compute the nonlinear optical properties of selected zinc-based MOF crystals, we predict polarization-entangled pair generation rates of $\sim 10^3-10^6$ s$^{-1}$mW$^{-1}$mm$^{-1}$ at 1064 nm, which are comparable with industry materials used in quantum optics. We find that the biaxial MOF crystal Zn(4-pyridylacrylate)$_2$ improves two-fold the conversion efficiency over a periodically-poled KTP waveguide of identical dimensions. This work underscores the great potential of MOF single crystals as entangled light sources for applications in quantum communication and sensing.
\end{abstract}

\date{\today}

\maketitle

\section{Introduction}

The genuine quantum correlations in the electromagnetic field are harnessed in optical quantum technology \cite{OBrien2009} for secure quantum communication \cite{Kimble2008,Wehner2018,Gomez2021}, quantum sensing \cite{Pirandola2018,Walborn2018} and quantum information processing \cite{Tasca2011,Wang2020}, are ultimately generated through the interaction of classical light with matter. The experimental performance of some quantum optical protocols can often be related to the properties of the optical materials used in implementations.  For instance, for communication protocols where photon entanglement is critical such as  device-independent quantum key distribution \cite{Primaatmaja2023,Zapatero:2023}, the secure bit rate can be related to the number of entangled photon pairs generated by a laser-driven nonlinear optical crystal \cite{Xu2013,Masanes2011,Liu2022diqkd}. 



Organic crystals have been studied extensively as candidate nonlinear optical materials due to their intrinsically high nonlinear optical response enabled by electron conjugation \cite{Eaton1991}. However, due to the relatively weak non-covalent nature of the interactions that drive molecular packing in organic crystals \cite{Li2009}, the optical stability of organic optical devices is typically lower than inorganic optical materials such as beta-barium borate (BBO) \cite{nikogosyan1991} or lithium niobate (LiNbO$_3$) \cite{schlarb1993}. In recent years, metal-organic framework (MOF) materials \cite{Furukawa2013,Colon2014} have emerged as competitive candidates for nonlinear optics due to their hybrid functional structure that combines the structural and optical stability of inorganic salts with the large nonlinearities of organic molecules \cite{Wang2012,Mingabudinova2016,Medishetty2017,Yang2018}. Experimental demonstrations of large second-order \cite{Cleuvenbergen2016-ZIF-8-SHG,Zhang2019-impressive,Chen2020-GIANT} and third-order optical nonlinearities \cite{Liu2016a} with polycrystalline MOF samples have stimulated materials science developments \cite{Chi-Duran2018,Enriquez2019,Garcia2020,duran2022} that have recently enabled the demonstration of three-wave and four-wave mixing processes with well-defined phase matching in millimeter-scale MOF single crystals \cite{Hidalgo-Rojas2023}. 

\begin{figure}[b]
\centering
\includegraphics[width=0.48\textwidth]{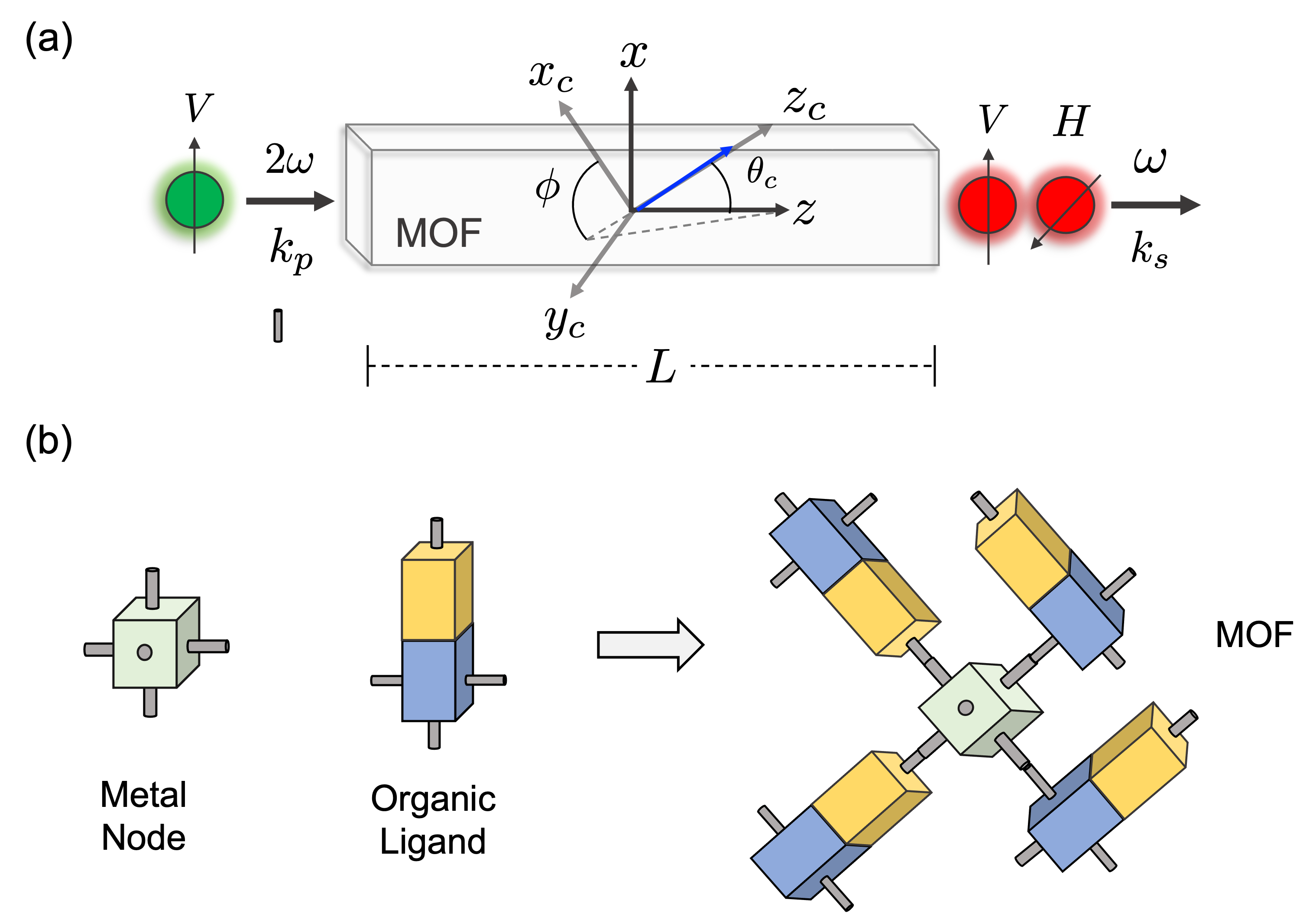}
\caption{(a) Illustration of degenerate collinear type-II SPDC with a single crystal MOF waveguide with length $L$. Pump photons with wavevector $k_p$ and frequency $2\omega$ enter the waveguide with lab-frame vertical polarization $V$ and produce a pair of polarization entangled photons with vertical and horizonal (H) polarizations in the same input direction, each with frequency $\omega$ and wavevector $k_s$. The crystal optic axis $z_c$ sets a polar angle $\theta$ and azimuthal angle $\phi$ with respect to the propagation direction $k$. (b) Schematics of the coordination of organic ligands with suitable metal ion nodes to form non-centrosymmetric MOF crystal structures. } 
\label{FIG.Diagram}
\end{figure}

These experimental breakthroughs increase the feasibility of using biaxial MOF single crystals for implementing quantum optical nonlinear processes such as spontaneous parametric down-conversion (SPDC). The generation of photon pairs having energy-time entanglement via type-I SPDC was studied theoretically for zinc-based MOF crystals with tetrazolate ligands \cite{ruben2021}, using a multi-scale modeling methodology that combines solid-state electronic structure calculations with phenomenological quantum optics theory. The influence of the structural and compositional details of tetrazole-based MOF crystals  on the nonlinear efficiency at room temperature was then computationally studied \cite{sanoj2023}, improving our understanding of the design features that are critical for discovering high-performing MOF nonlinear optical devices. 

The same theoretical methodology was recently used to identify a few dozen MOF materials that would be suitable entangled photon sources in collinear degenerate type-I SPDC \cite{Raj2023screening}, by screening a publicly available database containing about $10^5$ MOFs with known synthetic procedures \cite{Groom:bm5086}. In the present work, we further expand the theoretical scope of MOF crystals as entangled photon sources by characterizing the coincidence rate and correlation times of polarization-entangled photon pairs produced via collinear degenerate type-II SPDC in a MOF single crystal. The setup is  illustrated in Fig. \ref{FIG.Diagram}. We focus on selected zinc-based MOF structures of current experimental interest. To compare with standard entangled photon sources based on inorganic nonlinear crystals such as periodically-poled potassium titanyl phosphate (PPKTP) \cite{Schneeloch_2019}, we focus the analysis on single-mode MOF waveguides with sub-wavelength transverse dimensions and millimeter-scale propagation length.

In Sec. \ref{sec:phase matching}, we introduce the MOF crystals studied in this work and discuss the predicted phase matching conditions for collinear type-II SPDC at near-infrared wavelengths, extending the scope of the methodology developed previously \cite{ruben2021,Raj2023screening}. In Sec. \ref{sec:pair properties}, we discuss the generation rates of polarization-entangled photon pairs predicted for each MOF crystal and their intrinsic pair correlation times. Comparisons with experimental results for PPKTP waveguides of equivalent dimensions are made. In Sec. \ref{sec:conclusions}, we conclude and suggest future research directions. 

\section{Type-II Phase Matching with MOF crystals}
\label{sec:phase matching}

\begin{figure}[t]

\includegraphics[width=0.48\textwidth]{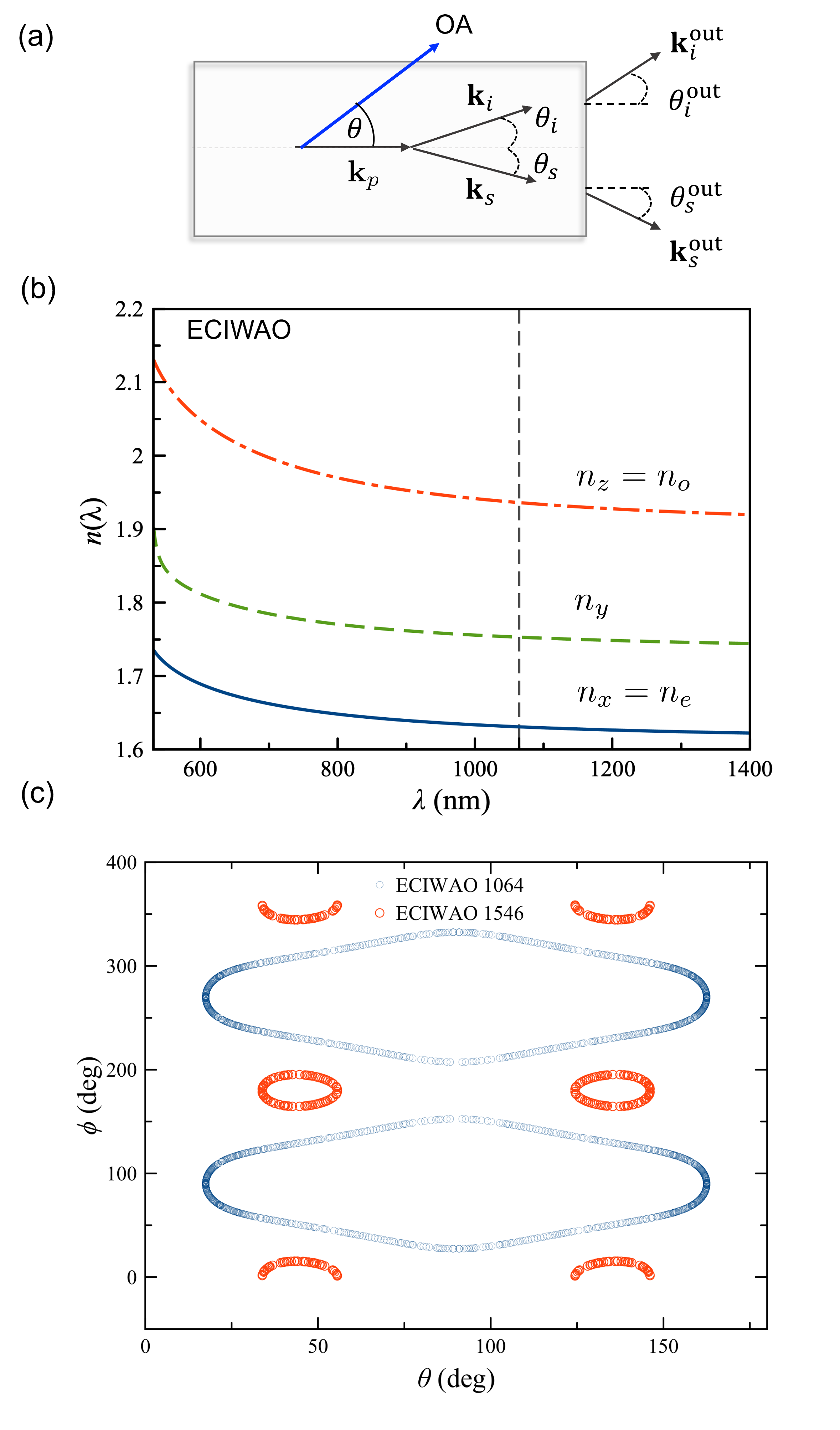}
\caption{(a) Ordinary and extraordinary output ray directions $\mathbf{k}_o^{\rm out}$ and $\mathbf{k}_e^{\rm out}$ for eeo type-II phase matching in the plane of the pump ray $\mathbf{k}_p$ and the crystal optic axis (OA). (b) Predicted Sellmeier curves for biaxial MOF  Zn(4-pyridylacrylate)$_2$ (ECIWAO). (c) Predicted tuning curves for ECIWAO as a function of the polar and azimuthal angles $\theta$ and $\phi$ at 1064 nm (blue circles) and 1546 nm (red circles) signal wavelengths.}
\label{Fig:Phase_matching_diagram}
\end{figure}

\subsection{Collinear Type-II SPDC with Biaxial Crystals}
Spontaneous parametric down-conversion (SPDC) \cite{Shih1988,Shih2003,Kiess1993} is a nonlinear optical process in which a pump photon of frequency $\omega_p$ and wavevector $\mathbf{k}_p$ is spontaneously converted into a pair of photons with frequencies and wavevectors $(\omega_1,\mathbf{k}_1)$ and $(\omega_2,\mathbf{k}_2)$, upon propagation through an optical medium with non-vanishing second-order optical susceptibility $\chi^{(2)}$ \cite{Schneeloch_2019}. Energy conservation requires that $\omega_1 + \omega_2 = \omega_p$ and momentum conservation implies the phase matching condition $\mathbf{k}_p=\mathbf{k}_1+\mathbf{k}_2$. The waves involved in the mixing process have a specific polarization configuration that determines the type of phase matching implemented.

\begin{table*}[thb]
\centering
\begin{tabular}{c c c  c  c  c  c  c}
\hline
\hline
Crystal & Axis & Type  & A & $B_1$ & $C_1$ & $B_2$ & $C_2$ \\
\hline
		&	$n_z$	& &	2.1072&  1.4606 & 1 4.804[4] & 9.5309[-2] & 1.1700[5]\\

MIRO-102 	& $n_y$ & negative  &  2.1385 &  1.1527 &  4.7765[4] &  7.6600[-2] &  1.1666[5]\\
	& $n_x$  &  &1.7890 &  5.030[-2] &  1.0701[5] &  7.432[-1] &  3.584[4]\\
\hline

& $n_z$ & &  2.8003 & 4.2754[-3] & 2.588765[5] & 1.6691 & 1.196108[5]\\
AQOROP 	&$n_y$ & negative &	2.5488 & 1.1271[-2] & 2.615202[5] & 9.2632[-1]  & 9.59809[4]\\	
		&$n_x$ & & 	2.0158 & 0.2528 & 8.56394[4] & 1.91085[-3]   & 2.643108[5]\\		
		
\hline

 & $n_z$ & &  1.866 & 5.370[-1] & 1.00393[5] & 1.2706   & 4.07423[4]\\
MOFTIL	& $n_y$ & negative &	1.8717 & 0.38553 & 1.05729[5] &  1.2577 & 3.8445[4]\\	
		&$n_x$ & 	& 1.9196[1] & 9.6197[-2]& 7.91035[4] & -1.7001[1] & -7.3178[2]\\		
\hline

 & $n_z$ & &  2.2698 & 2.37285[-2] & 2.487748[5] & 1.42196 & 1.046126[5]\\
ECIWAO	&$n_y$ & positive & 2.2368 & 0.75972 & 8.981561[4] & 7.18826[-3]   & 2.7285289[5]\\	
		&$n_x$ & &	1.91004& 5.6766[-1] & 8.756205[4] & 1.31190[-2]   & 2.446305[5]\\				
\hline
\hline

\end{tabular} 
\caption{Ordinary ($n_o$) and extraordinary ($n_e$) Sellmeier coefficients for the MOF crystals studied in this work, obtained by fitting ab-initio calculations to Eq. (\ref{eq:sellmeier}). The notation $a \cross 10^b \equiv a [b]$ is used.}
\label{tab:sellmeier}
\end{table*}

Under conditions of type-II phase matching, the polarization state of one of the output photons is parallel to that of the pump and the second photon in the output pair has orthogonal polarization \cite{Boyd-book,zhang2017}. Denoting the plane that contains the pump ray $\mathbf{k}_p$ and crystal optic axis (OA) as the extraordinary direction (e) and the axis orthogonal to it as the ordinary direction (o), in a type-II SPDC process the polarization combinations ($oeo,ooe,eoe,eeo$) are allowed for the waves at $(\omega_p,\,\omega_1\,\omega_2)$, respectively. The relative strength of these wave-mixing channels is given by the combined influence of the orientation of the pump rate and polarization relative to the incidence face of the crystal and the contraction of the nonlinear susceptibility tensor with the input and output polarization states. The latter determines the effective nonlinear coefficient \cite{Boyd-book}
\begin{equation}\label{eq:deff}
d_{\rm eff}\equiv \chi_{ijk}^{(2)}E_i(\omega_p)E_j(\omega_1)E_k(\omega_2), 
\end{equation}
where $E_i(\omega)$ denotes the electric field component of the wave at frequency $\omega$ in the cartesian crystal frame ($i=x_c,y_c,z_c$). We follow standard notation and set the optic axis as the $z_c$ direction (see Fig. \ref{FIG.Diagram}a). Given the polar angle $\theta$ of optic axis relative to the pump ray and the azimuthal angle $\phi$, the vectorial phase matching equation $\Delta \mathbf{k}\equiv \mathbf{k}_p-\mathbf{k}_1-\mathbf{k}_2=0$ implies the following condition for collinear degenerate SPDC in a biaxial crystal \cite{yao1984calculations}
\begin{eqnarray} 
\frac{1}{2} (n_{\omega,+}+n_{\omega,-}) = n_{2\omega,-} 
\label{eq:PM biaxial}
\end{eqnarray}
where the left-hand side contains the propagation index of the signal and idler photons and the right-hand side refers to the pump field. For biaxial crystals, the effective index $n_w$ at frequency $\omega$ for a given propagation direction $\mathbf{k}=(k_x,k_y,k_z)$ can be obtained for each wave by solving the quadratic equation in $x= 1/n_\omega^2$ \cite{yao1984calculations,ito1975generalized,huo2015effective}
\begin{eqnarray} 
\frac{k_x^2}{x-n_{x,\omega}^{-2}} +\frac{k_y^2}{x-n_{y,\omega}^{-2}} + \frac{k_z^2}{x-n_{z,\omega}^{-2}} = 0
\label{eq:biaxial phase matching}
\end{eqnarray}
where $k_x = \rm{sin}\theta \,\rm{cos}\phi$, $k_y = \rm{sin}\theta \,\rm{sin}\phi$ and $k_z = \rm{cos}\phi$. $n_{x,\omega}$, $n_{y,\omega}$ and $n_{z,\omega}$ are the crystal indices along orthogonal directions. The solution reads
\begin{eqnarray} 
n_{\omega,\pm} = \frac{\sqrt{2}}{\sqrt{-B_\omega \pm \sqrt{B_\omega^2-4C_\omega}}}
\label{eq:n_omega}
\end{eqnarray}
where the sub-index $\pm$ refers to the fastest ($+$) and the slowest ($-$) refractive indices. The solution is given in terms of $
B_\omega = [-k_x^2 (b_\omega+c_\omega)-k_y^2 (a_\omega+c_\omega)-k_z^2 (a_\omega+b_\omega)]$, $C_\omega = [k_x^2 b_\omega c_\omega + k_y^2 a_\omega c_\omega + k_z^2 a_\omega b_\omega]$, where $
a_\omega = n_{x,\omega}^{-2}$ , $b_\omega = n_{y,\omega}^{-2}$, and $c_\omega = n_{x,\omega}^{-2}$.


Biaxial crystals can be negative or positive, this depends on a relationship between the refractive indices given by  $n_z - n_y > n_y - n_x$ for positive biaxial crystals and $n_y-n_x > n_z - n_y$ for negative biaxial crystals \cite{Raj2023screening,yao1984calculations}. This is particularly relevant when computing the crystal $d_{\rm eff}$ because if it is positive the polarization configuration is ooe, while if it is negative the configuration is eoe.

\subsection{Birefringent Metal-Organic Frameworks}

We follow the procedure in Ref. \cite{ruben2021} to compute the refractive indices $n_z$, $n_y$ and $n_x$ as a function of wavelength $\lambda$ for the non-centrosymmetric zinc-based MOF crystals listed in Table \ref{tab:sellmeier}. The crystal structures of MIRO-103 contains cadmium nodes \cite{Enriquez2019}.  MOF crystals AQOROP, MOFTIL and ECIWAO were identified in Ref. \cite{Raj2023screening} as suitable candidates for optical frequency conversion. Further details of the chemical structure and crystal properties of the MOFs studied in this work can be found in the  Supplementary Material (SM).  

Table \ref{tab:sellmeier} shows the Sellmeier coefficients of the MOF crystals, obtained by fitting the calculated dielectric tensor elements $\epsilon_{ii}$  to the two-term expression \cite{Ghosh1997}
\begin{eqnarray}
n^{2}(\lambda) = A + \frac{B_1\lambda^{2}}{\lambda^{2}-C_1} + \frac{B_2\lambda^{2}}{\lambda^{2}-C_2}.
\label{eq:sellmeier}
\end{eqnarray}
The diagonal tensor elements $\epsilon_{ii}$ were calculated using periodic density-functional theory (DFT), as described in  \cite{ruben2021,sanoj2023,Raj2023screening}. Figure \ref{Fig:Phase_matching_diagram}b shows the three Sellmeier curves used for MOF  Zn(4-pyridylacrylate)$_2$ (ECIWAO), highlighted here as an example. We set $n_z = \sqrt{\epsilon_{zz}}$, $n_y = \sqrt{\epsilon_{yy}}$ and $n_x = \sqrt{\epsilon_{xx}}$ for this crystal. 

Figure \ref{Fig:Phase_matching_diagram}c shows the tuning curves for ECIWAO that correspond to perfect phase matching for collinear degenerate ooe type-II SPDC (positive biaxial) at 1064 nm. For each combination of $\theta$ and $\phi$ that gives phase matching, we can calculate $d_{\rm eff}$ and search for the angular variables that optimize the optical nonlinearity $d_{\rm eff}$. 

\begin{figure}[t]
\includegraphics[width=0.47\textwidth]{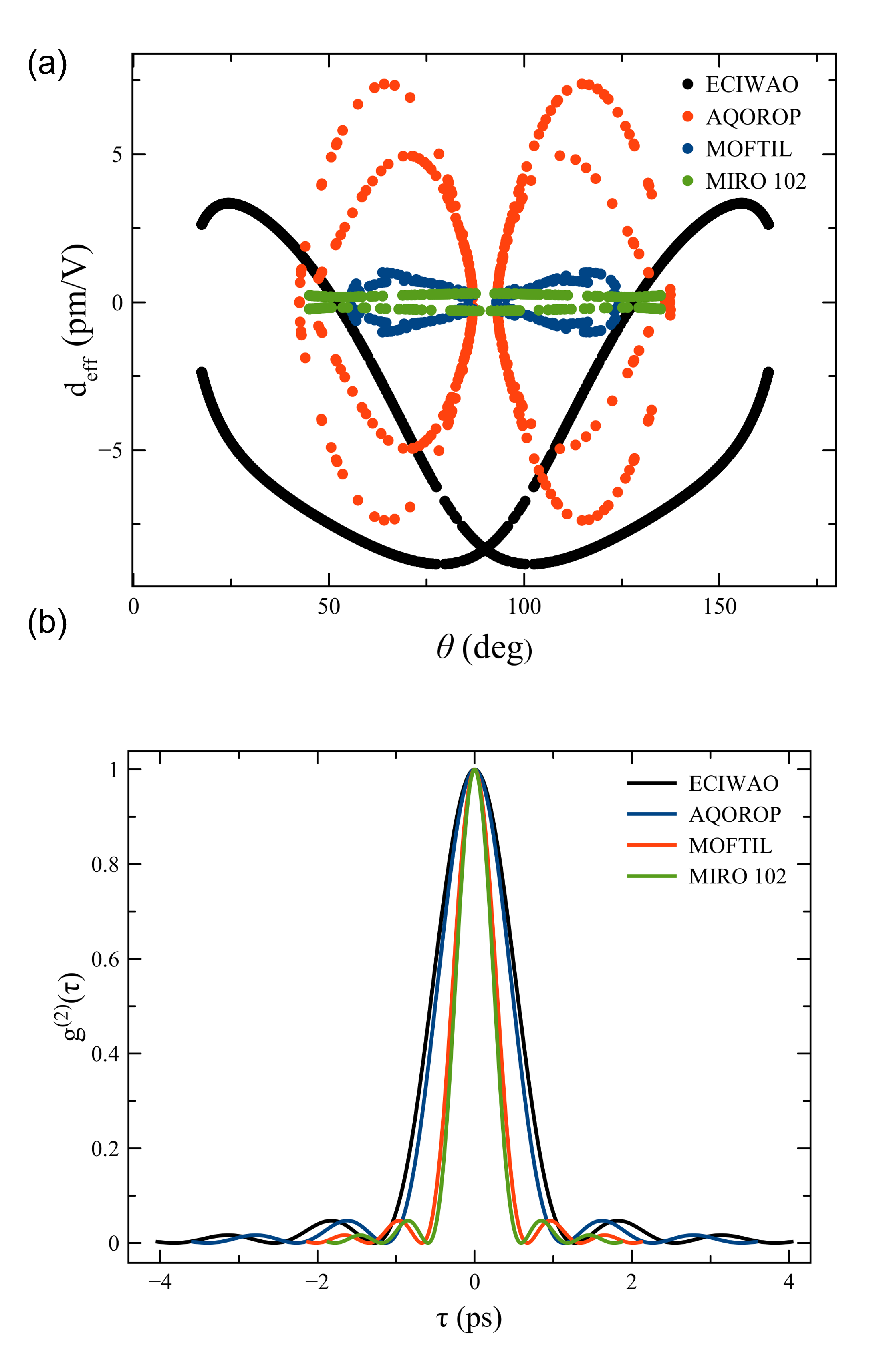}

\caption{(a) Effective nonlinearity $d_{\rm eff}$ in pm/V, as a function of polar angle $\mathrm{\theta}$ for selected MOF structures. The azimuthal angle $\phi$ is fixed to enable collinear type-II phase matching at 1064 nm. (b) Glauber intensity correlation function $g^{(2)}(\tau)$ a function of delay time $\tau$.} 
\label{fig:g2}
\end{figure}

$d_{\rm eff}$ is obtained for each MOF  by contracting the calculated second-order susceptibility tensor elements $\chi^{(2)}_{ijk}$ with the unit polarization vectors of the pump and signal waves involved in the type-II SPDC process, according to Eq. (\ref{eq:deff}). The {\it ab-initio} tensor elements are taken from Refs. \cite{ruben2021,Raj2023screening}. Figure \ref{fig:g2}a shows  $d_{\rm eff}$ in units of pm/V as a function of $\theta$ given an optimal choice of $\phi$, for selected MOFs. The variation of the maximum value of $d_{\rm eff}$ at the optimal pair angles $\theta$ and $\phi$ spans two orders of magnitude from $0.29$ pm/V for MIRO-102 to about $8.8$ pm/V for ECIWAO.  $d_{\rm eff}$ curves for additional MOF crystals can be found in the SM. These nonlinearities should be compared with typical $d_{\rm eff}$ values obtained for commercial-grade crystals such as KDP (0.38 pm/V), BBO (1.94 pm/V), LiNbO$_3$ (4.7 pm/V), and KTP (3.5 pm/V), measured in second-harmonic generation (SHG) \cite{Eckardt1991}. Strongly nonlinear MOF crystals such as ECIWAO, MOFTIL and AQOROP can thus potentially outperform the frequency-conversion efficiency of conventional nonlinear optical crystals.

\section{Polarization-Entangled Photon Pairs}
\label{sec:pair properties}

Having established the frequency dependence of the phase mismatch $\Delta \mathbf k$ and obtained the effective nonlinearity $d_{\rm eff}$ of a target MOF crystal using {\it ab-initio} electronic structure calculations, we now use these material parameters to construct a quantum optical description of SPDC that enables an analysis of the temporal correlations of the generated photon pair \cite{ruben2021,Raj2023screening} as well as the overall brightness of the entangled photon source in terms of the number of pairs produced per unit time, normalized by pump power and  propagation length.

\begin{table*}[t]
\begin{tabular}{l  c 	c  c c  c	  c	c}
\hline
\hline
Crystal & $\lambda_s$ (nm) &  $R\,(\mathrm{s}^{-1}\mathrm{mW}^{-1}\mathrm{mm}^{-1})$\hspace*{4mm} & $d_{\rm eff}$ (pmV$^{-1})$ &$\tau_{L}\,(\mathrm{fs})$ & $E_G$ $({\rm eV})$ & $\theta$\,({\rm deg}) & $\phi$\, (deg)\\
\hline
MIRO-101 & 1064	 & $ 7.6 \times 10^4$  & 0.86 & 480.28  & 3.15 & 28.9 & 305.2\\
MIRO-102 & 1064	& $ 5.6 \times 10^3$   & 0.29 & 527.7 & 3.10 & 91.5 & 340.7\\
MIRO-103 &1064 	& $ 3.1 \times 10^3$ & 0.13 &407.05 & 2.98 & 36.5 & 29.9 \\
MOFTIL &	 1064	& $ 6.6 \times 10^4$ & 1.01 & 599.25 & 3.26 & 116.7 & 162.2 \\
		& 1546	&  $ 8.0 \times 10^4$	& 0.38	& 288.2	& 3.26		& 109.9 & 169.4 \\
AQOROP & 1064	& $1.9 \times 10^6$ & 7.38 & 1009.72 & 2.36 & 114.8 & 19.0 \\
 		& 1546 	&$ 1.5 \times 10^5$ & 0.56 & 338.6 & 2.36 &  56.1 & 7.4 \\
ECIWAO & 1064	& $ 3.4 \times 10^6$ & 8.85 & 1136.1	& 2.28 & 77.5 & 329.0\\
 	     &	1546 	& $3.4 \times 10^4$ & 0.67 & 310.45 & 2.28  & 42.2 & 344.8\\
PPKTP (Exp. \cite{Schneeloch_2019}) & 1546 & $1.67\pm 0.04\times 10^6$ & $3.18 \pm 0.32$ &833.91  & 3.52 & - & -\\
\hline
\hline
\end{tabular} 
\caption{Number of entangled photon pairs ($R$) generated via SPDC at signal wavelength $\lambda_s$ per unit second, per milliwatt of pump power, per millimeter crystal length, for the single-mode MOF waveguides studied in this work. The intrinsic two-photon correlation time $\tau_L$, crystal band gap $E_G$, optic axis polar angle $\theta$ and optimal azimuthal angle $\phi$ for collinear biaxial phase matching are also given. Where MIRO-101 and MIRO-102 are uniaxial crystals, for more information see SM. The experimental pair generation rates and correlation times obtained at 1546 nm using periodically-poled potassium titanyl phosphate (PPKTP) waveguides \cite{Schneeloch_2019} are also shown for comparison.}
\label{tab:generation rates}
\end{table*}

\subsection{Two-photon correlation times}

The spectral and temporal properties of entangled photon pairs produced via SPDC are described by a two-photon wavefunction that for a monochromatic pump and one-dimensional propagation can be written as \cite{Shih2003} 
\begin{equation}
\ket{\Psi} = A  \int d\omega_s \Phi(\Omega_p,\omega_s) \hat a^\dagger_s(\Omega_s)\hat a^\dagger_i(\omega_p-\omega_s)\ket{0},
\label{eq:two-photon}
\end{equation}
where $\hat a^\dagger_s$ and $\hat a^\dagger_i$ are the field creation operators for the {\it signal} and {\it idler} photons in the pair, respectively. $\Omega_p$ is the fixed pump frequency and $\omega_s$ is the frequency of     one of the photons in the pair (e.g., signal). The other photon frequency (idler) is fixed by energy conservation as $\omega_i=\Omega_p-\omega_s$. As discussed below, the constant $A$ encodes the dependence of the two-photon state with physical parameters such as $d_{\rm eff}$, $L$, and pump power. For the one-dimensional case discussed here, $A$ also includes the transverse field overlap of the three waves involved \cite{Schneeloch_2019}. Generalizations of Eq. (\ref{eq:two-photon}) for other types of phase matching conditions involving three-dimensional field propagation can be found in Refs. \cite{Boucher2015,Fiorentino2007,Hutter2021,Walborn_2012,Walborn2004,Burkalov2001}.

The  joint spectral amplitude $\Phi$ in Eq. (\ref{eq:two-photon}) is determined by the details of phase mismatch configuration along the propagation direction. For type-II phase matching we have $\Phi(\omega)= {\rm sinc}(\omega DL/2)\,{\rm exp}[i D L/2] $, where 
\begin{equation} 
D = \frac{dk_{o}}{d\omega_{o}}\Bigg|_{\Omega_{o}}-  \frac{dk_{e}}{d\omega_{e}}\Bigg|_{\Omega_{e}}
\label{eq:group_diff}
\end{equation}
is the inverse group velocity difference at the central output frequencies $\Omega_{o}$ and $\Omega_{e}$, corresponding to the ordinary and the extraordinary waves. For degenerate phase matching we have $\Omega_o=\Omega_e=\Omega_p/2$. $k_o$ and $k_e$ are the ordinary and extraordinary wavenumbers at these frequencies. The wavenumber derivatives can be obtained numerically using the Sellmeier parameters in Table \ref{tab:sellmeier}.

The probability of two photons to arrive at equally distant photodetectors with delay time $\tau$ is given by the two-time intensity correlation function $G^{(2)}(\tau)=|\langle 0|\hat E_1(\tau)\hat E_2(0)|\Psi\rangle|^2$ \cite{Boucher2015,Schneeloch_2019,Shih2003}, where $\hat E_i(t)$ represents the effective field operator at the detector location. Normalizing to the zero-delay value $G^{(2)}(0)$ and inserting Eq. (\ref{eq:two-photon}) gives the intensity autocorrelation function 
\begin{equation} 
g^{(2)}(\tau) = \left| \int \,d\mathrm{\nu} \,{\rm sinc}(\nu DL/2) e^{-\left(\frac{\mathrm{\nu}^2}{\sigma^2}\right)} e^{-i\nu\tau}\right|^2
\label{eq:g2}
\end{equation}
where the integral is evaluated numerically over a finite range of detuning $\nu$ around the signal frequency $\Omega_p/2\pm \nu_{\rm max}$. The frequency cutoff is chosen  proportional to the characteristic frequency $\nu_L=(DL/2)^{-1}$, which sets the bandwidth of the joint spectral amplitude. $\sigma$ is the detector bandwidth.

Figure \ref{fig:g2}b shows the $g^{(2)}$ functions predicted by Eq. (\ref{eq:g2}) for the MOFs in panel \ref{fig:g2}a. The detector bandwidth is $\sigma = 1\,{\rm nm}$ and the crystal length $L = 1\,{\rm mm}$. The theoretical autocorrelation times broadly vary in the range $\sim 0.5-2.0$ ps (FWHM), depending primarily on the birefringence of the crystal and the slope of the Sellmeier curves at perfect phase matching. The correlation times for MOF crystals are therefore equivalent to those available using conventional nonlinear optical crystals \cite{Shih2003}, with ECIWAO and AQOROP having the largest autocorrelation times in the broadband detection scheme here considered.  

\subsection{Photon pair generation rates}

In order to calculate the absolute pair generation rate via SPDC, the two-photon wavefunction $\ket{\Psi}$ in Eq. (\ref{eq:two-photon}) needs to be fully characterized in terms of physical parameters. The joint spectral amplitude $\Phi(\omega)$ is determined by the phase matching conditions. An explicit expression for the constant prefactor $A$ can be obtained from microscopic derivations based in perturbation theory, assuming three-dimensional wave propagation in the crystal \cite{Boucher2015,Fiorentino2007,Hutter2021,Schneeloch_2019,Walborn_2012,Walborn2004,Shih2003,Burkalov2001}. 

We follow closely the derivation in Ref. \cite{Schneeloch_2019} and assume propagation of a monochromatic collimated Gaussian pump beam of wavelength $\lambda_p$ on a rectangular MOF waveguide with a sub-wavelength transverse area $L_x\times L_y$ and propagation length $L_z\equiv L$ in the millimeter regime (i.e., $L\gg 2\lambda_p$). The waveguide is assumed to support a single guided mode and the transverse intensity profiles of the pump and down-converted light is taken as  zeroth-order Hermite-Gaussian modes with characteristic widths $w_p$ (pump), $w_s$ (signal), and $w_i$ (idler). 
For simplicity, we set signal and idler transverse widths equal ($w_s = w_i$). The counting rate for collinear type-II SPDC is thus

\begin{eqnarray} 
R &=&  \frac{|E_{p}^{0}|^{2}(d_{\rm eff})^{2}L^{2}}{2\pi c^{2}} \frac{n_{gs}n_{gi}}{n_{s}n_{i}}\left|\frac{w_{p}^{2}}{w_{s}^{2}+2w_{p}^{2}}\right|^{2}\nonumber\\ 
&& \times \int d\omega\,\omega\,(\Omega_p - \omega)\, \mathrm{sinc}^{2} (\Delta k(\omega) L/2)
\label{eq:counting_rate}
\end{eqnarray}
where $|E^0_p| =|D_p^0|/{e_o n^2}$ is proportional to the monochromatic pump peak magnitude $|D_p^0|$.  The power delivered by a Gaussian pump beam is given by $P = c |D_p^0|^2 \pi \sigma_p^2 /{n^3 \epsilon_0} $. $n_s$ and $n_i$ are the refractive indices of signal and idler fields, respectively.  $n_{gs}$ and $n_{gi}$ are the corresponding group indices. The frequency-dependent phase mismatch $\Delta k(\omega)$ is obtained from the Sellmeier curves as discussed above, and $d_{\rm eff}$ is evaluated at the optimal azimuthal angle (see Fig. \ref{fig:g2}a) for each MOF. 

Table \ref{tab:generation rates} lists the photon pair generation rates $R$ predicted for the MOF crystals studied in this work. For direct comparison with previous measurements in rectangular single-mode PPKTP waveguides \cite{Schneeloch_2019}, we set $w_s = 1.875 \,\mu{\rm m}$ and $w_p = 0.875 \,\mu{\rm m}$. The band gap $E_G$, polar and azimuthal angles that gives optimal collinear phase matching, and the corresponding optimal  $d_{\rm eff}$, are also shown.

Tetrazole-based MOF waveguides (MIRO-102) are found to be less efficient than PPKTP at generating entangled photon pairs by up to three orders of magnitude. Similarly, the predicted photon conversion rate of MOFTIL crystals ($\approx 6.6\times 10^4 \,{\rm s}^{-1}{\rm mW}^{-1}\mathrm{mm}^{-1})$ is 25 times smaller than the measured brightness of PPKTP \cite{Schneeloch_2019}. On the other hand, AQOROP and ECIWAO crystals at 1064 nm exhibit roughly the same or twice the pair generation rates than PPKTP at 1546, respectively. However, the relatively low band gaps of these high-performing nonlinear materials ($\sim2.3 $ eV) do not necessarily prevent absorptive losses at the pump wavelength 532 nm, suggesting that these MOF crystals could be better suited for down-conversion at longer pump wavelengths ($\lambda_p>800 nm$). Note that at longer signal wavelengths, the magnitude of the $d_{\rm eff}$ tensor is expected to decrease due to dispersion \cite{zhang2017}. For completeness, Table \ref{tab:generation rates} also shows the calculated pair generation rates for SPDC and other relevant system parameters at 1536 nm.


\section{Conclusions and Outlook}
\label{sec:conclusions}

We explored non-centrosymmetric MOF single crystals as suitable quantum light source materials. We theoretically studied the features that characterize the indistinguishability and brightness of type-II SPDC entangled photon sources and identified specific MOF structures with known synthetic procedures that have the potential to overcome the theoretical down-conversion efficiency limits of widely used nonlinear optical crystals such as PPKTP, developed for decades for optimizing efficiency   \cite{Zhong2012,Jachura2014,Cozzolino2019,Gomez2022}. We showed that one-dimensional waveguides of MOF Zn(4-pyridylacrylate)$_2$ [ECIWAO, Table \ref{tab:generation rates}] of millimeter lengths can generate $3.3 \times 10^6$ entangled photon pairs per second, per mW of pump power, per millimeter (s-mW-mm), a two-fold improvement over the measured conversion efficiency of PPKTP waveguides with equal dimensions \cite{Schneeloch_2019}. Tetrazole-based MOF crystals such as  MIRO-102, for which growth protocols to the millimeter regime are known \cite{Garcia2020}, are shown to have sub-optimal conversion efficiencies ($5.6\times 10^4$ pairs per s-mW-mm) relative to PPKTP but compare very well with respect to other commonly used optical crystals such as KDP ($\sim 30$ pairs per s-mW-mm \cite{Mosley2008}).

High-quality entangled photon sources are enabling tools in photonic quantum technology \cite{OBrien2009,Wang2020}. Modern applications in quantum communication strongly rely on SPDC sources for high-dimensional photonic entanglement and heralded single photons \cite{Carine:2020}, making the prospects for applicability of efficient nonlinear optical materials such as MOFs in next-generation quantum devices highly promising. Although recent breakthroughs in MOF crystal engineering  have enabled the fabrication of bulk-size crystals that are suitable for free-space quantum optics \cite{Garcia2020,Hidalgo-Rojas2023}, other quantum applications such as photon triplet generation via third-order SPDC \cite{Corona2011,Corona2011pra,Moebius2016,Cavanna2020} are also promising research directions because they do not require millimeter-scale single crystals and could be implemented using polycrystalline samples. In addition to a better understanding of their growth mechanism \cite{Colon2019}, the heat transport and optical stability properties of the non-porous MOF materials discussed in this work need to be further explored in order to fully assess the potential of this novel optical material class for applications in quantum optics.


\section{Acknowledgements}
We thank Esteban Sepúlveda, Gustavo Lima, Stephen Walborn and Dinesh Singh for discussions. S.P., R.A.F. and F.H. thank support from ANID through grants FONDECYT Regular No. 1221420 and the Millennium Science Initiative Program ICN17\_012. R.A.F thank financial support by ANID Fondecyt Postdoctoral 3220857. S.R. and Y.J.C. acknowledge computational resources from the Center for Research Computing (CRC) at Notre Dame.

\bibliography{typeII_mof.bib}

\end{document}


\onecolumngrid
\renewcommand\theequation{S.\arabic{equation}}
\renewcommand\thefigure{S.\arabic{figure}}   
\section{Theoretical framework}
\subsection{Effective non-linearity}

The electric field component in the ordinary axis is given by $E_j^o(\omega)$ = $(a_j)E^o(\omega)$ and the electric field component in the 
extraordinary axis is given by $E_j^e(\omega)$ = $(b_j)E^e(\omega)$, where,
\begin{equation}
    (a_j) = 
    \begin{pmatrix}
       \sin \phi \\ -\cos \phi \\ 0 
    \end{pmatrix},
    \quad
    (b_j) =
    \begin{pmatrix}
        -\cos \theta \cos \phi \\ -\cos \theta \sin \phi \\ \sin \theta
    \end{pmatrix}.
\end{equation}
where $a_j$ and $b_j$ correspond to the vector of ordinary and extraordinary polarization, respectively. For type-II negative uniaxial, $d_{\text{eff}}$ is calculated as
\begin{equation}
     P_{eo}^e (\omega_3) = b_i d_{ijk}(\omega_3,\omega_2,\omega_1) b_j a_k E_j(\omega_2) E_k(\omega_1),
     \label{negative}
\end{equation}
and for a positive uniaxial crystal, we use  
\begin{equation}
     P_{eo}^o (\omega_3) = a_i d_{ijk}(\omega_3,\omega_2,\omega_1) b_j a_k E_j(\omega_2) E_k(\omega_1),
     \label{positive}
\end{equation}
$P(\omega_3)$ is the dielectric polarization of the crystal at the pump frequency $\omega_3 = \omega_1 + \omega_2$ ($\omega_3 = 
2\omega$ and $\omega_2$ = $\omega_1$ = $\omega$). A conventional $e$-$eo$ $type-II$ condition was used  for negative 
uniaxial crystal, where the entangled signal and idler are polarized along the ordinary axis ($o$) and extraordinary axis and the pump is polarized along 
the extraordinary axis ($e$). For a positive uniaxial crystal, a conventional $o$-$eo$ type-II condition was used, where the entangled signal and idler are polarized along the ordinary ($o$) and extraordinary ($e$) axis, and the pump along the ordinary axis ($o$).

\subsection{Uniaxial treatment}

 Following standard notation and setting the optic axis as the $z_c$ direction. Given the polar angle $\theta_c$ of optic axis relative to the pump ray, the vectorial phase mismatch $\Delta \mathbf{k}\equiv \mathbf{k}_p-\mathbf{k}_1-\mathbf{k}_2$ vanishes when the following system of equations is satisfied 
\begin{eqnarray}
    {k}_{e} \rm {cos} \theta_e + {k}_{o} \rm {cos}\theta_o = {k}_p \label{eq:PM1}\\
    {k}_{e} \rm {sin} \theta_e = {k}_{o} \rm {sin} \theta_o, \label{eq:PM2}
\end{eqnarray}
where we arbitrarily label $\mathbf{k}_1\equiv \mathbf{k}_{\rm o}$ and $\mathbf{k}_2\equiv \mathbf{k}_{\rm e}$. The angles $\theta_e$ and $\theta_o$ are defined with respect to pump ray, which is taken along the lab-frame $z$-direction, as illustrated in Fig. \textcolor{blue}{1}a. The wavenumber  $k=n\omega/c$ is directly related to the frequency-dependent refractive index function $n(\omega)$, where $c$ is the speed of light.

The refractive indices along the ordinary ($n_o$) and extraordinary ($n_e$) directions are different for type-II crystals due to birrefringence. Given a dielectric tensor $\epsilon$ of the nonlinear crystal in a principal coordinate frame (i.e., $\epsilon_{ij}=\epsilon_{ii}\delta_{ij}$), we can obtain the indices $n_{ii}=\sqrt{\epsilon_{ii}}$ and solve for the direction of signal rays $\theta_o$ and $\theta_e$ in Eqs. (\ref{eq:PM1})-(\ref{eq:PM2}). From Snell's law, we can obtain the corresponding outside angles into air according to $\sin\theta_{e}^{o} = n_{\mathrm{eff}}(\lambda_e,\varphi_e)\sin \theta_e$ and $\sin\theta_{o}^{o} = [\lambda_o/\lambda_e]n_{\mathrm{eff}}(\lambda_e,\varphi_e)\sin \theta_e$, respectively. $n_{\rm eff}$ is the effective index experienced by the extraordinary ray at an angle $\varphi_e$ with respect to the optic axis at wavelength $\lambda_e$. For $\varphi_e=0$, the extraordinary wave index is $n_o$ and for $\varphi_e=\pi/2$ the index is $n_e$ \cite{Boyd-book}.  

The phase matching conditions in Eqs. (\ref{eq:PM1}) and (\ref{eq:PM2}) depend parametrically on the optic axis angle $\theta_c$ through the effective index $n_{\rm eff}$. Since the orientation of the optic axis relative to the lab-frame $z-$axis can in principle be chosen by conveniently cutting the crystal faces \cite{Dawson:1926}, we set $\theta_c$ as a free parameter that is chosen such that perfect phase matching is achieved in a collinear configuration, i.e., $\mathbf{k}_p=k_{p}\hat{\mathbf{z}}$ with $\theta_e^{\rm out}=\theta_o^{\rm out}=0$, for type-II SPDC from 532 nm to 1064 nm. This collinear configuration is relevant to describe SPDC in one-dimensional waveguides with sub-wavelength transverse dimensions \cite{Schneeloch_2019}.
\clearpage

\section{MOF CRYSTAL STRUCTURES}
\begin{figure}[!htbp]
\centering 
\includegraphics[width=.6\linewidth]{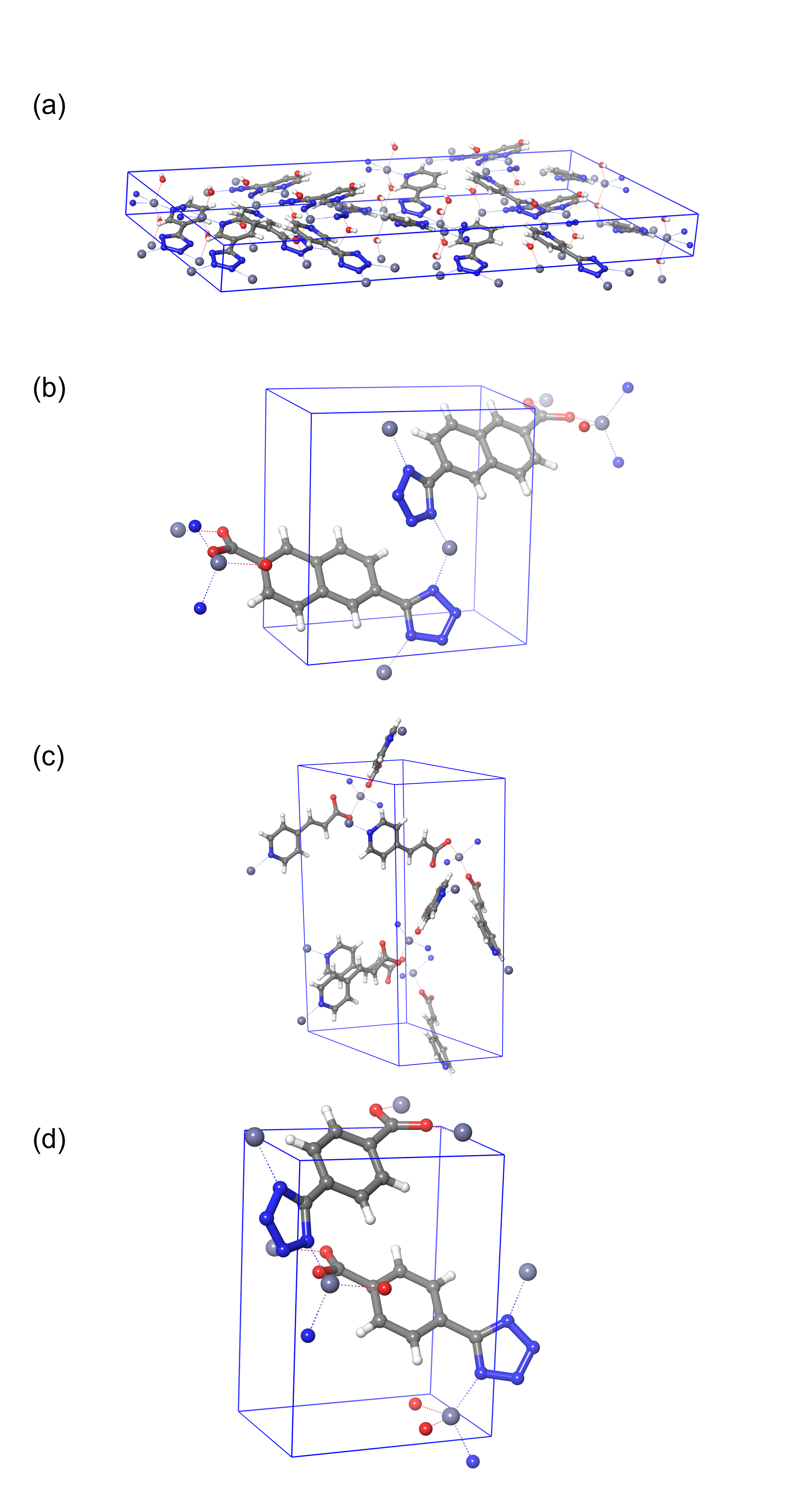}
\caption[Unit Cell of the MOFs]{Unit Cell of the MOFs: a) MIRO-102, b) AQOROP, c) ECIWAO and d) MOFTIL01. Hydrogen, Carbon, Oxygen, Nitrogen and Zinc atoms are colored: white, gray, red, blue and purple respectively. Axis are showed only in (c) for clarity and simplicity of the graph and are representative of the axis for every MOFs unit cell}
\label{fig:Unit_Cells}
\end{figure}

\clearpage
\begin{table*}[!htbp]
\centering
\begin{adjustbox}{width=\textwidth}
\begin{tabular}{cccccc}
\hline
MOF      &  MOFs Name     & Ligand name ( corrected )                           & Chemical formula (ligand) & Space Group  & Referensce \\
MIRO-101 &  $Zn(3\text{-}ptz)_{2}$           & 3-(pyridyl)tetrazol                   &  C6H5N5     & I42d         & \cite{MIRO-101} \\
MIRO-102 &  $$[Zn(OH)(3\text{-}ptz)]$$         & 3-(pyridyl)tetrazol                   &  C6H5N5     & Fdd2         & \cite{MIRO-102-103} \\
MIRO-103 &  $$[CdN3(3\text{-}ptz)]$$           & 3-(pyridyl)tetrazol                   &  C6H5N5     & P3221        & \cite{MIRO-102-103}  \\
ECIWAO   &  $$Zn(4\text{-}pyridylacrylate)2$$  & 4-pyridin-acrylate                    &  C8H6NO2${-}$  & Cc           &  \cite{ECIWAO} \\
AQOROP   &  $$[Zn(C12H6N4O2)]n$$        & 6-(1H-tetrazol-5-yl)-2-naphthoic acid &  C12H8N4O2  & Pc           &  \cite{AQOROP} \\
MOFTIL01 &  $$[Zn(tzba)]n$$             & (4-(tetrazol-5-yl)benzoat)            &  C8H5N4O2${-}$  & Pc           &  \cite{MOFTIL01} \\
\hline
\end{tabular}
\end{adjustbox}
\caption[Selected MOFs information: name, ligand and space group]{Selected MOFs information: name, ligand and space group.}
  \label{tab:MOFs_info}{}
\end{table*}

\clearpage
\section{ORIENTATIONS FOR PERFECT PHASE MATCHING}    
\begin{figure}[!htbp]
\centering
\includegraphics[width=.50\linewidth]{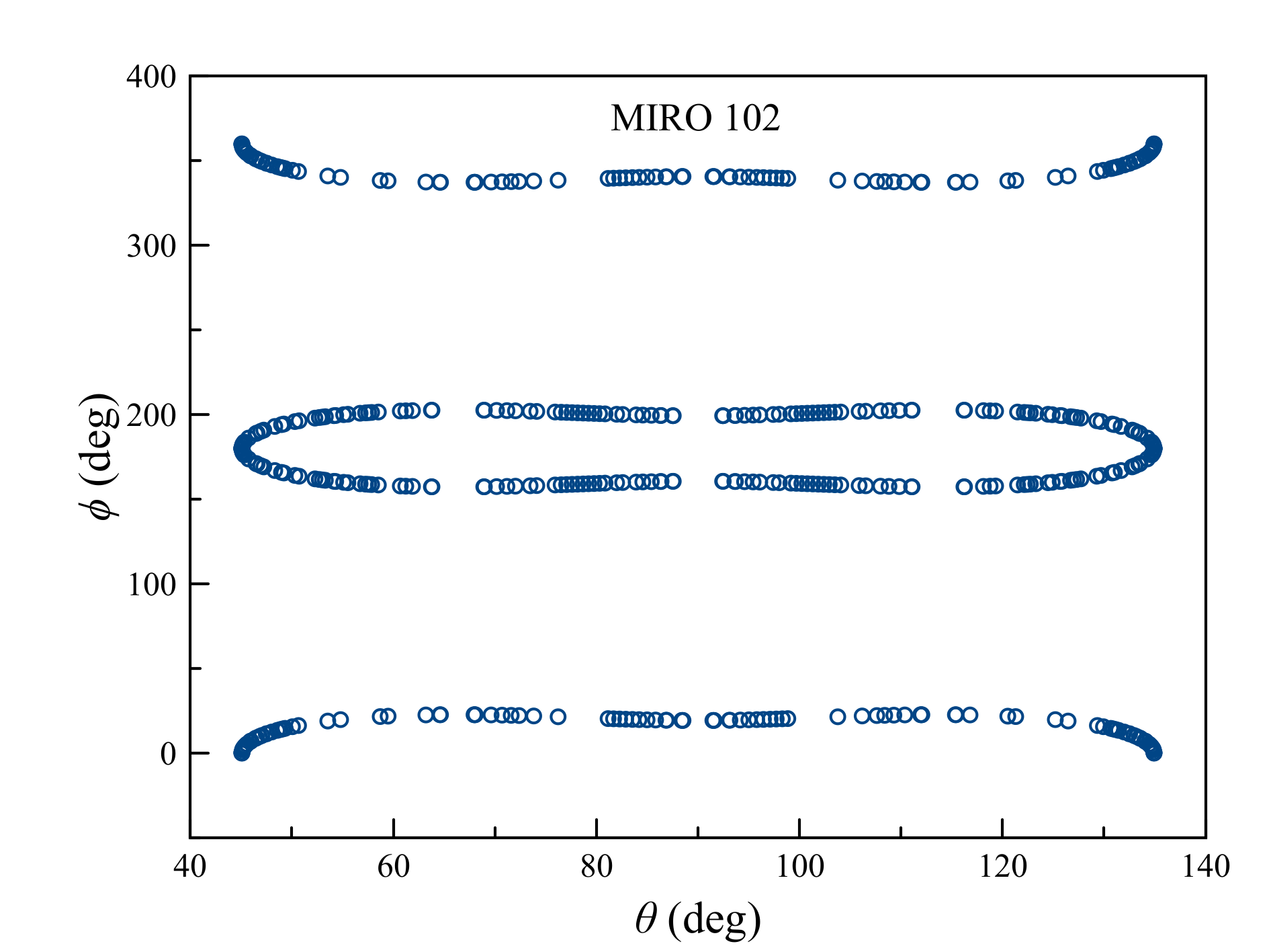} \\
\includegraphics[width=.50\linewidth]{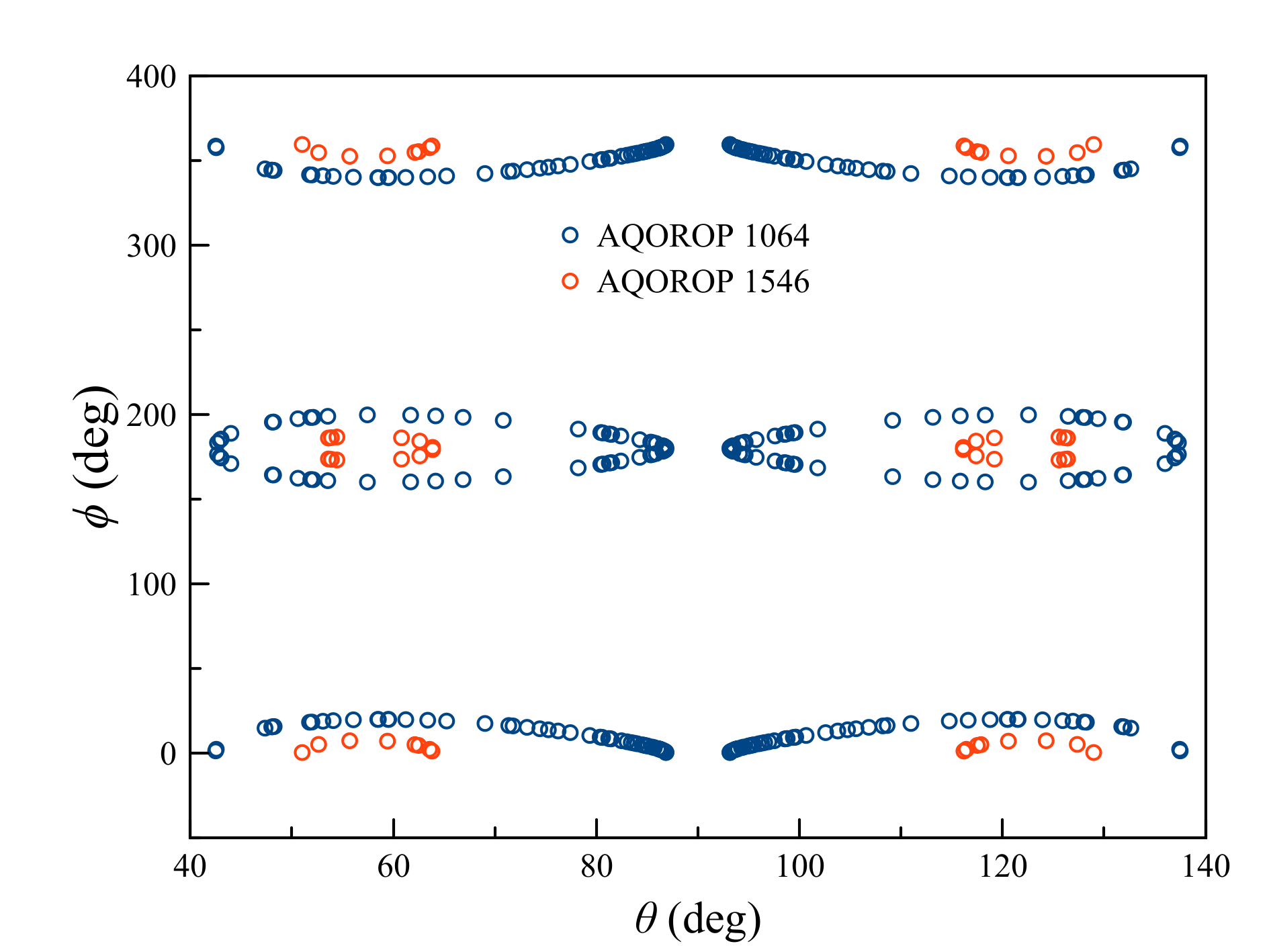} \\
\includegraphics[width=.50\linewidth]{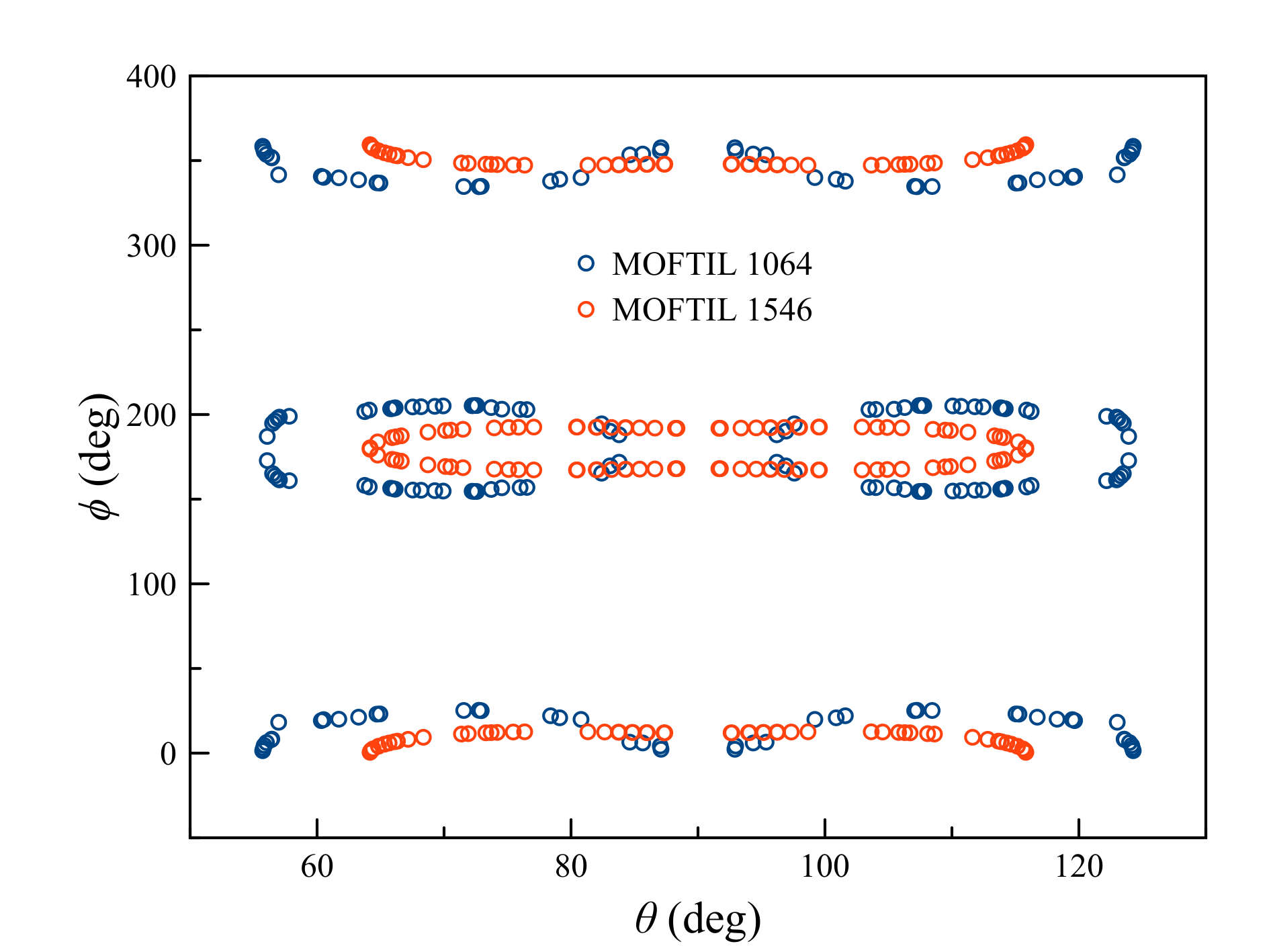}
 \caption[Tuning curves for selected MOF structures]{Tuning curves for selected biaxial MOF structures, the  blue dots correspond to a pumped wavelength of 1064 and the red dots correspondo to a pumped wavelength of 1546. Where each point on the plot ensure collinear non-degenerate type-II SPDC.}
\label{fig:false-color}
\end{figure}

\clearpage
\begin{figure}[!htbp]
\centering
\includegraphics[width=.75\linewidth]{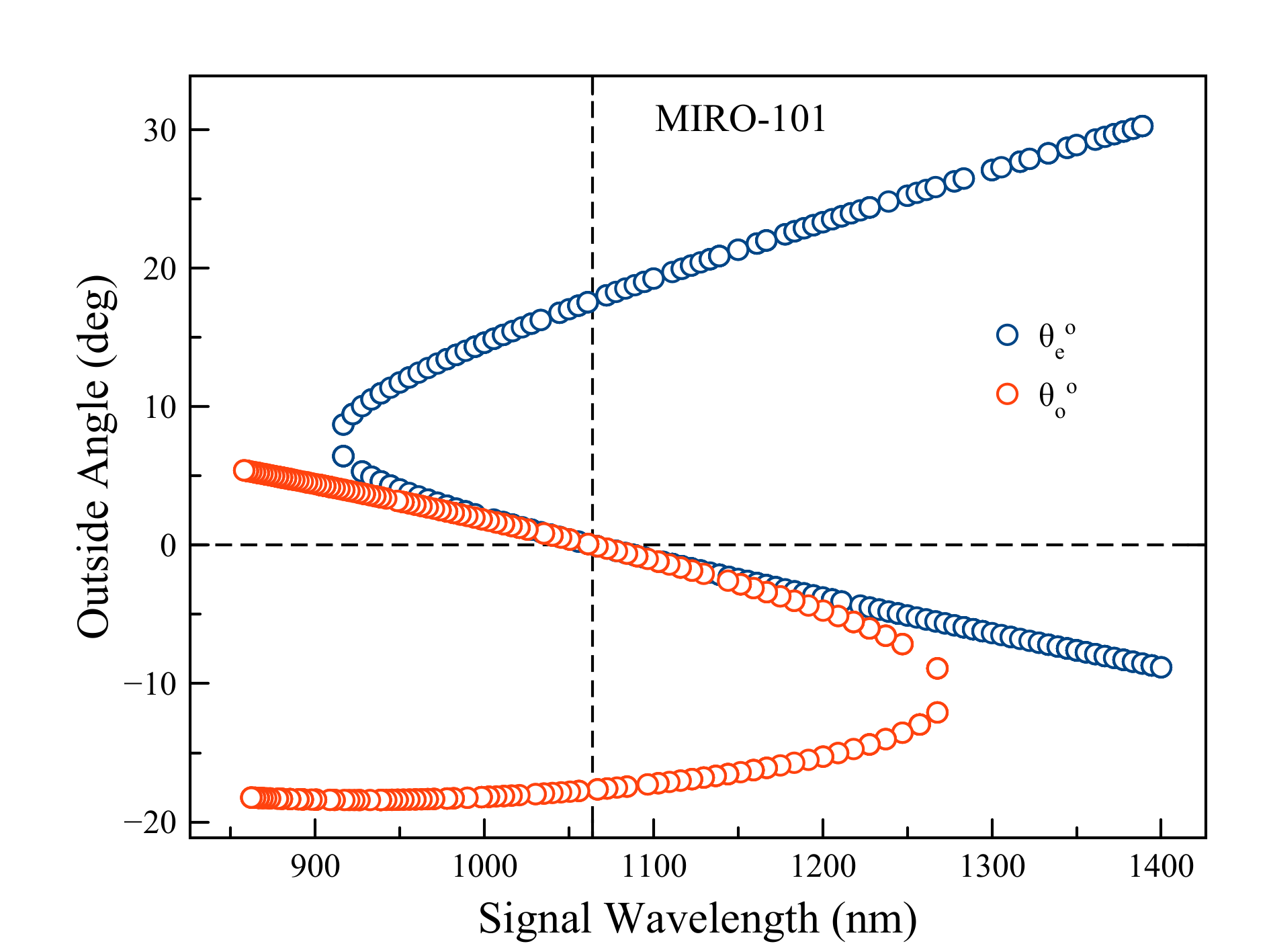} \\
\includegraphics[width=.75\linewidth]{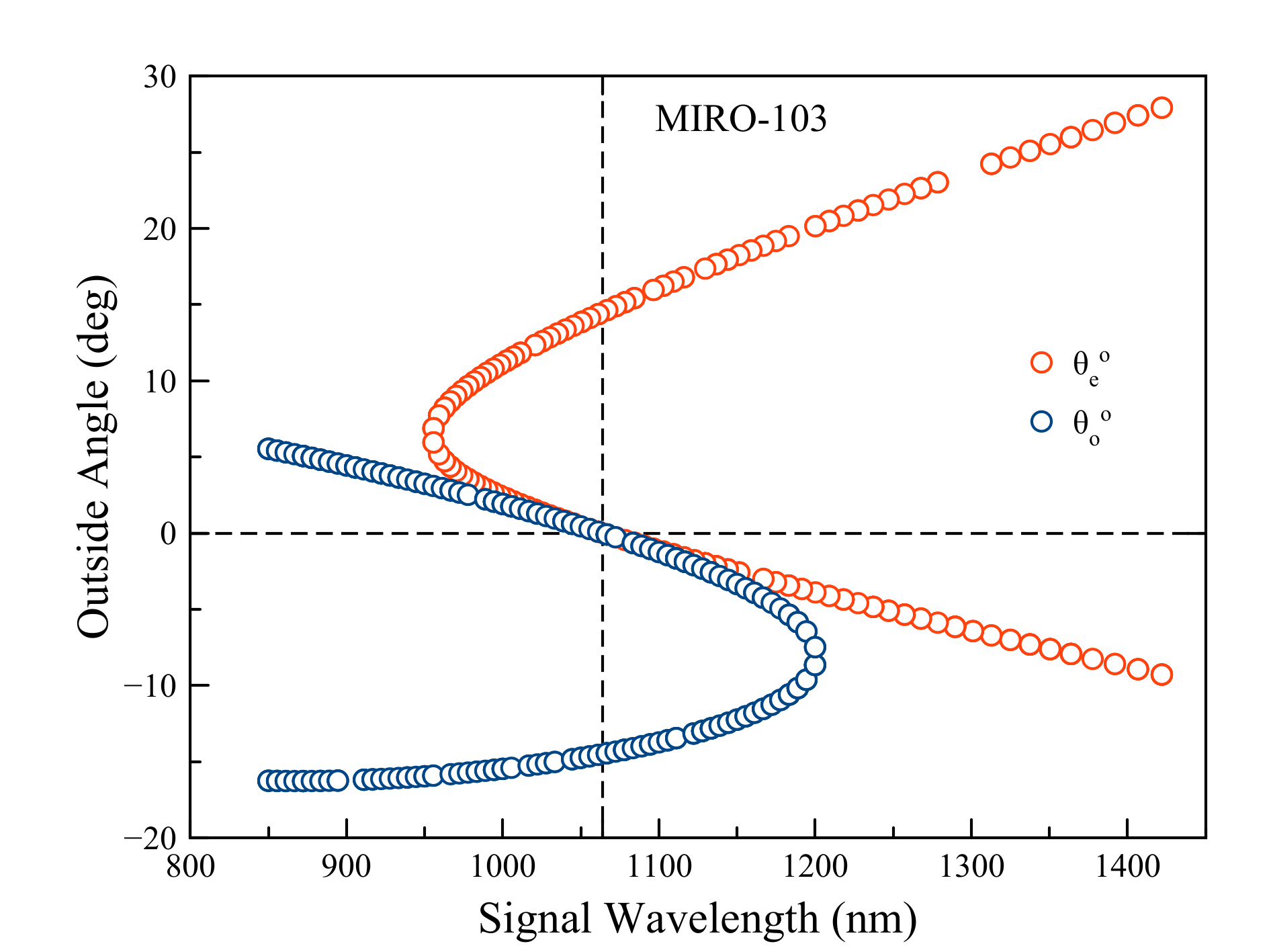}
\caption[Tuning curves for selected MOF structures]{Tuning curves for selected biaxial MOF structures, the  blue dots correspond to the outside angle of the ordinary polarized photon and the blue dots to the extraordinary polarized photon. For each tuning curve $\theta$ was fixed to ensure an outside angle of 0 for both photons at 1064 nm.}
\label{fig:false-color}
\end{figure}

\clearpage

\section{ORIENTATION DEPENDENCE OF EFFECTIVE NONLINEARITY} 

\begin{figure}[!htbp]
\centering
\includegraphics[width=.50\linewidth]{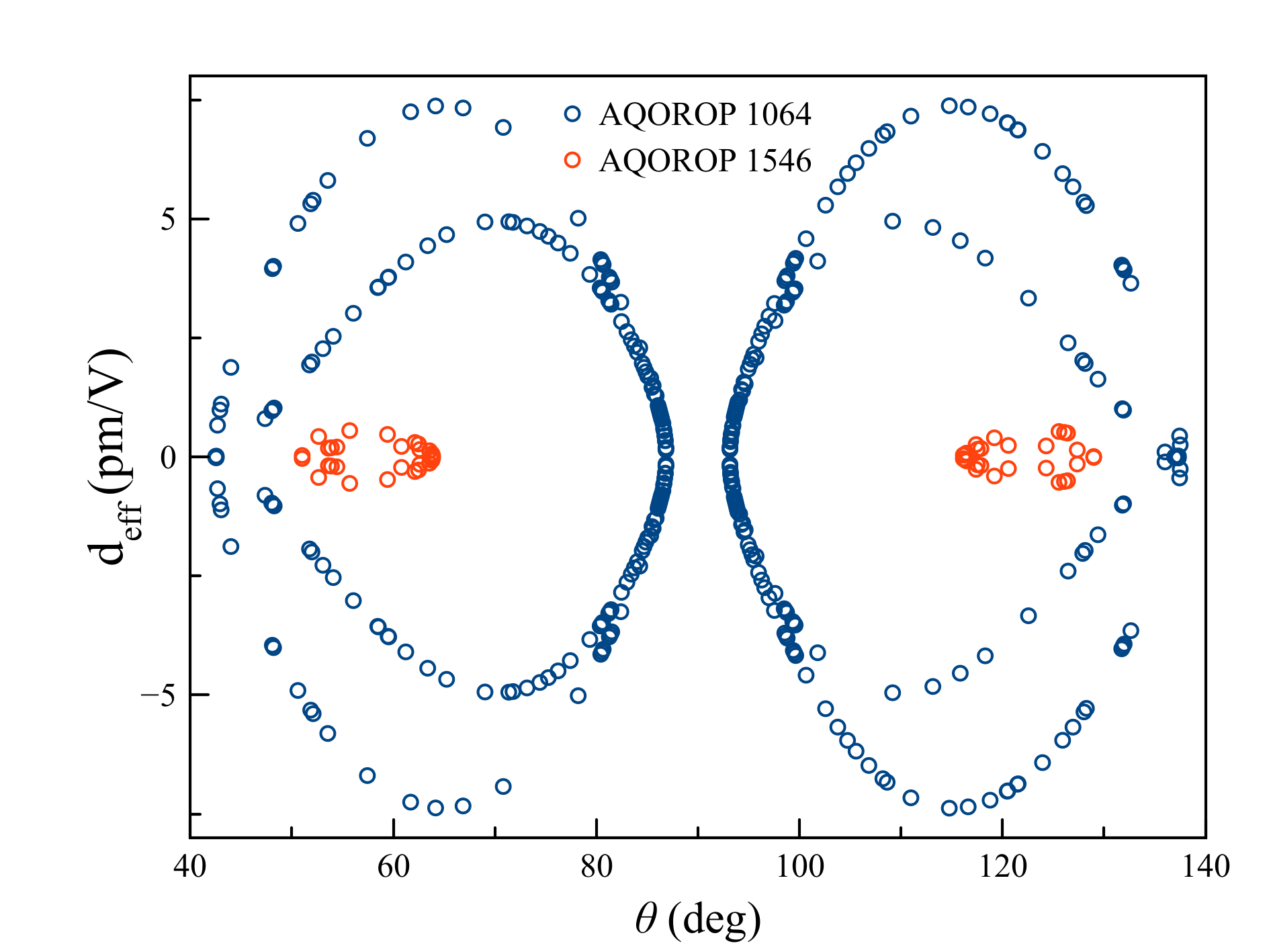}
\includegraphics[width=.50\linewidth]{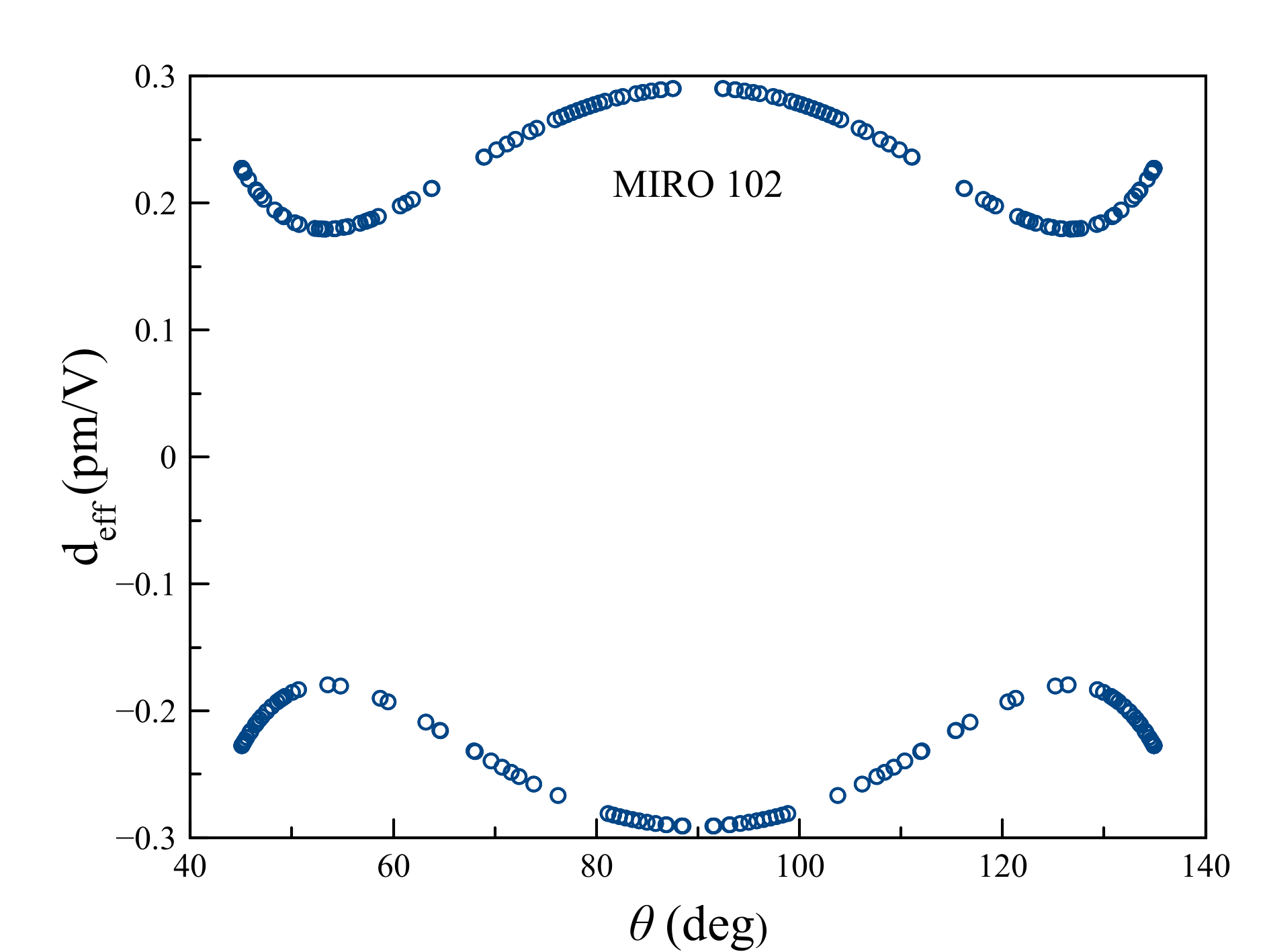}
\includegraphics[width=.50\linewidth]{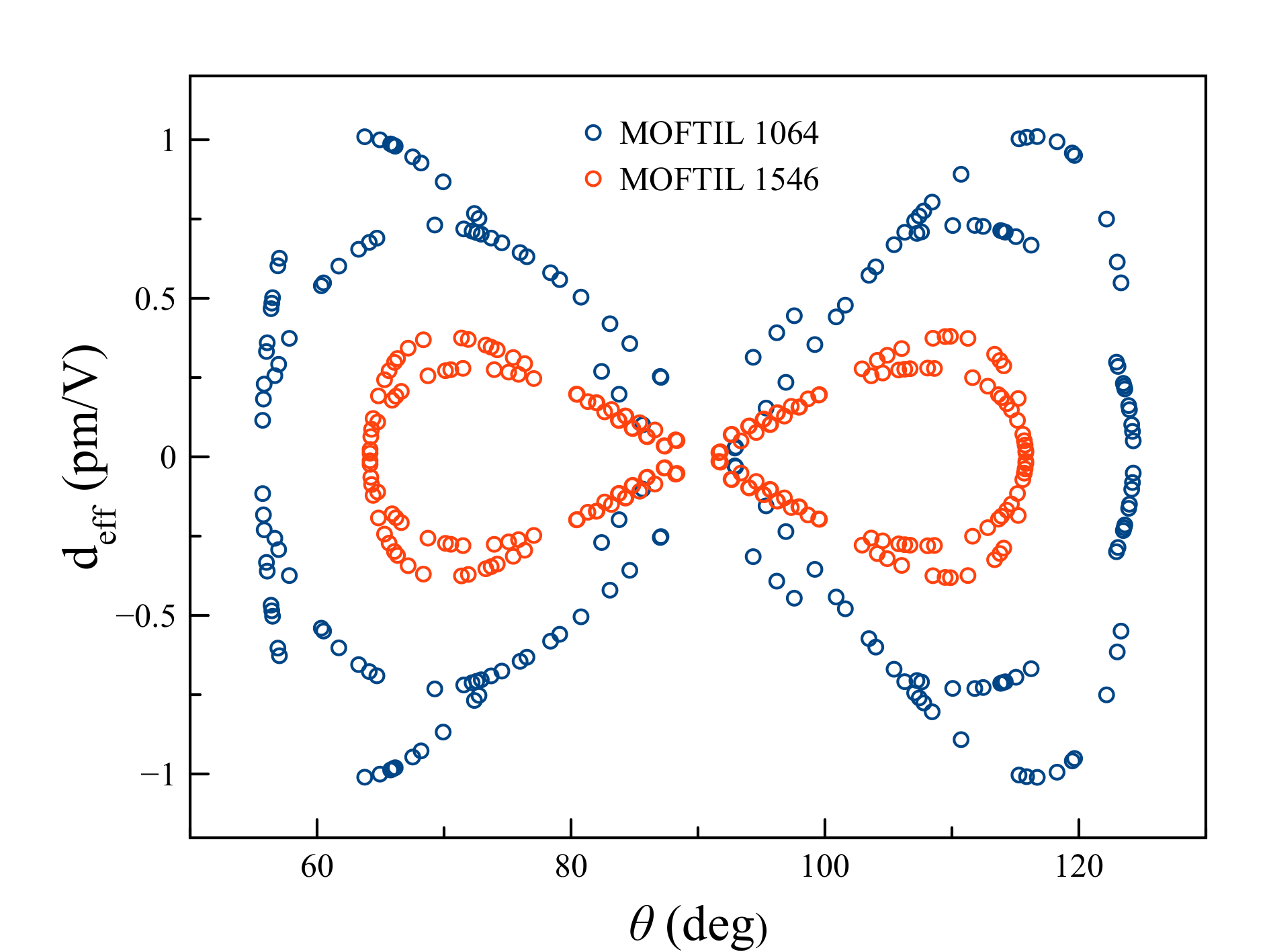}
\caption[Effective nonlinearity deff of selected MOFs]{Effective nonlinearity $d_{\rm eff}$ (pm/V), as a function
of the polar angle $\theta$ for selected MOF structures. For each crystal
the polar angle $\theta$ and the azimuthal angle $\phi$ 
form a pair that enable collinear type-II phase matching at 1064 nm and 1546 nm.}
\label{fig:Deff all MOFs}
\end{figure}

\clearpage
\begin{figure}[!htbp]
\centering
\includegraphics[width=.75\linewidth]{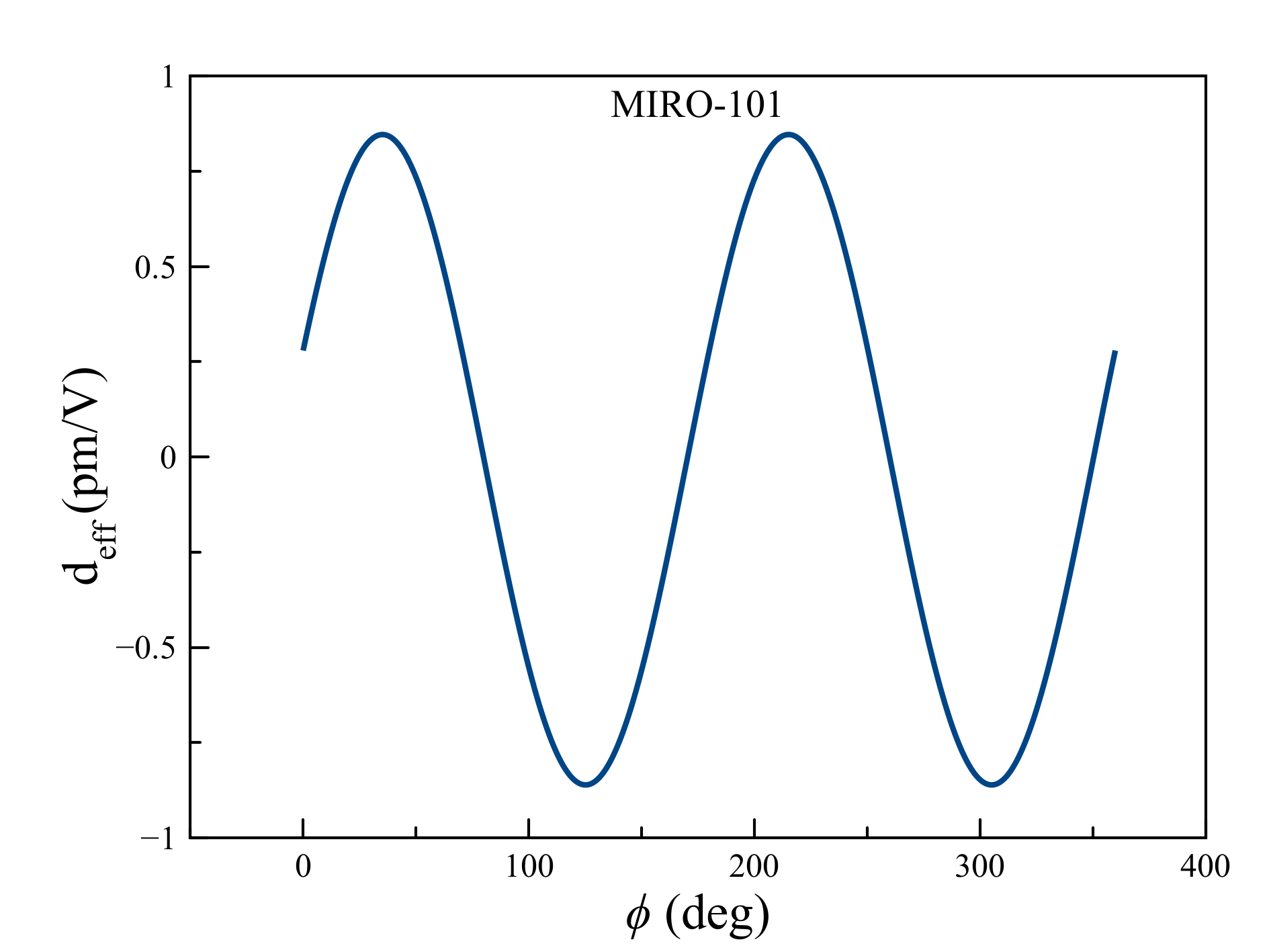}
\includegraphics[width=.75\linewidth]{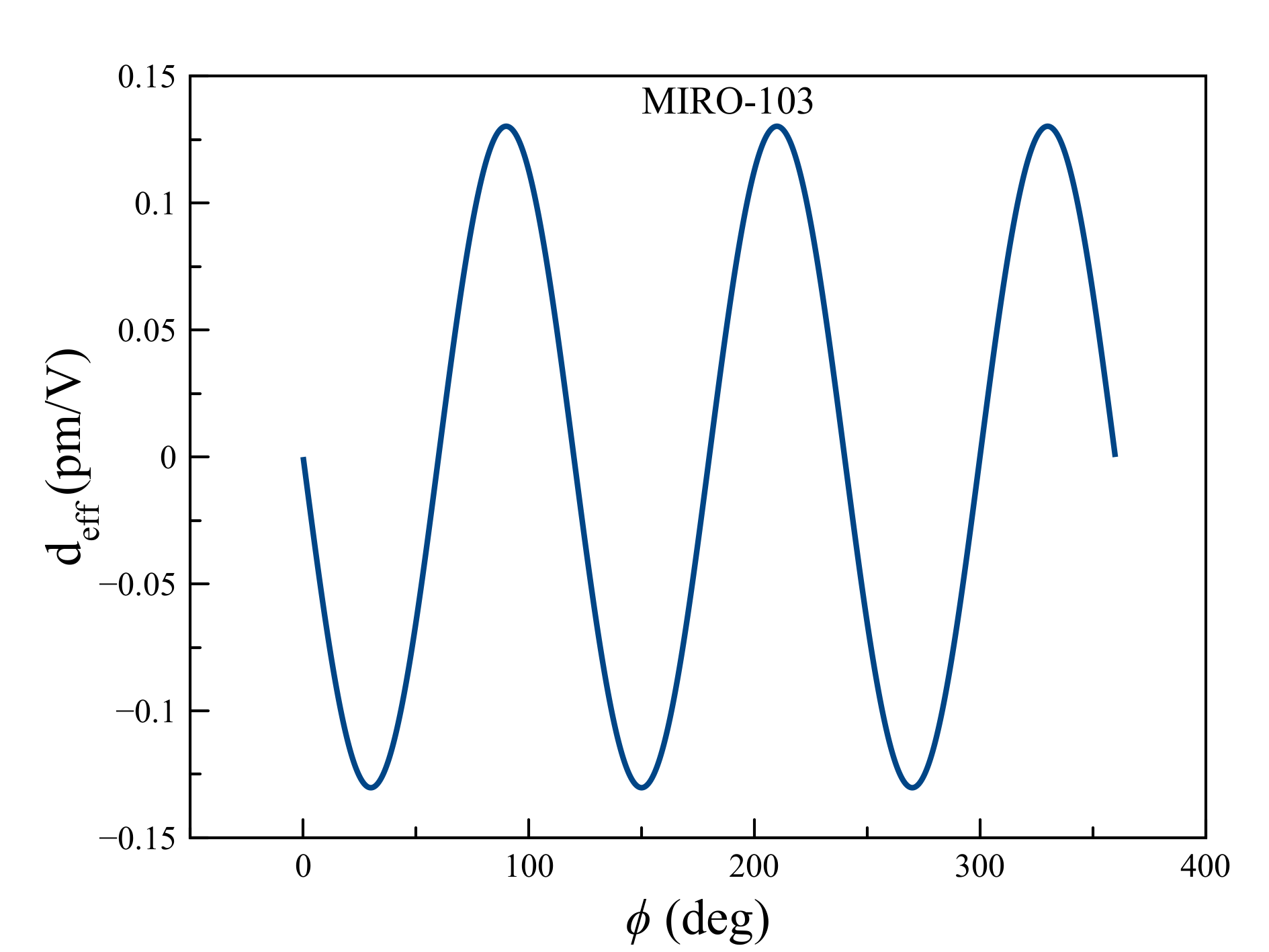}
\caption[Effective nonlinearity deff of selected MOFs]{Effective nonlinearity $d_{\rm eff}$ (pm/V), as a function
of azimuthal angle $\phi$ for selected biaxial MOF structures, where $\theta$ was fixed to ensure collinear SPDC.}
\label{fig:Deff all MOFs}
\end{figure}

\clearpage
\section{COMPARISON WITH CONVENTIONAL OPTICAL MATERIALS} 

\begin{figure}[!htbp]
\centering
\includegraphics[width=.70\linewidth]{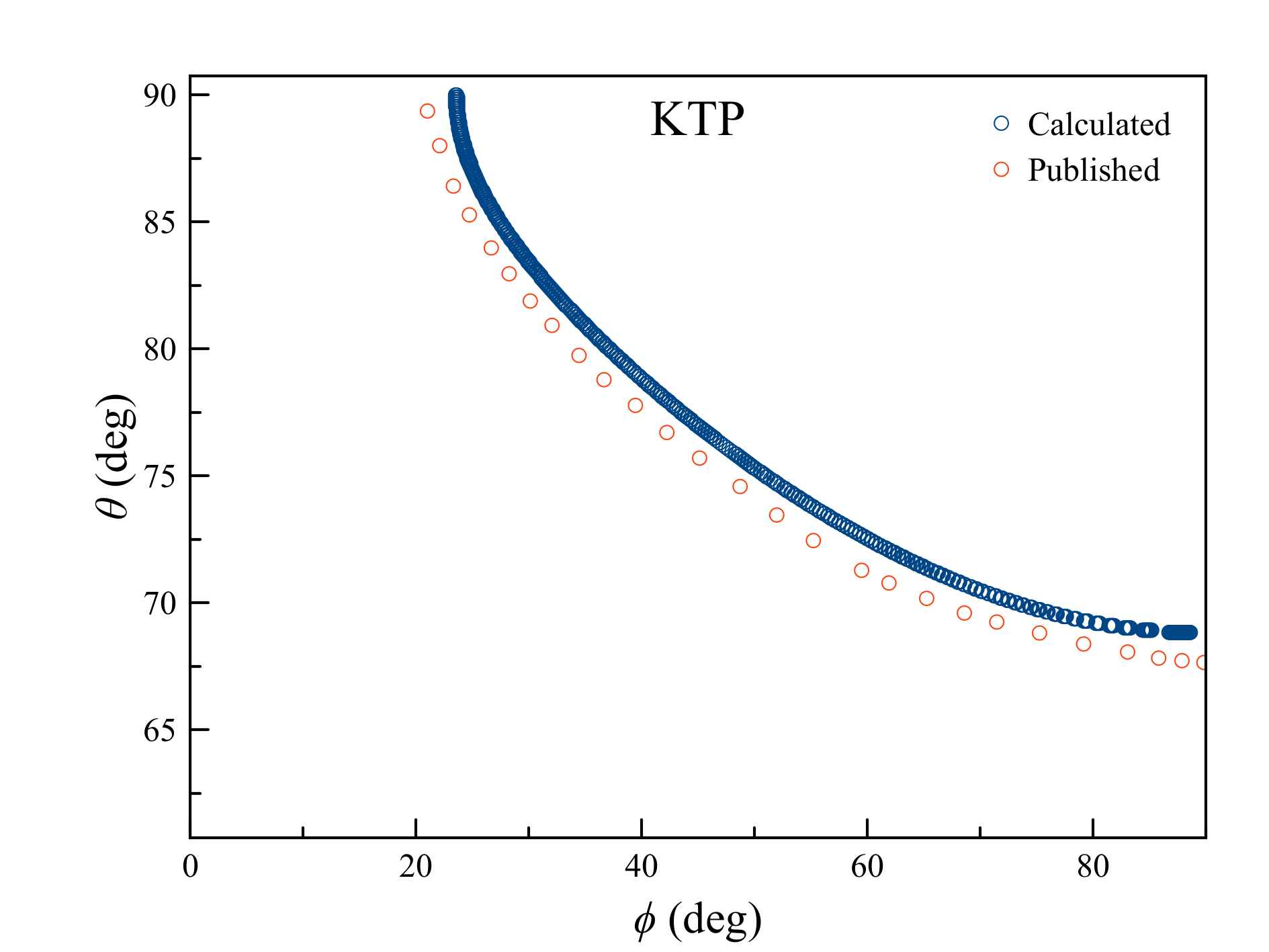}
\caption[KTP tuning curve]{Predicted tuning curves KTP at a pumped wavelenght of 532 nm, where $\theta$ and $\phi$ act as a pair that ensure collinear non-degenerate Type-II SPDC, where the blue dots are the calculated tuning curves and the red dots are the published values \cite{yao1984} for comparison.}
\label{fig:KTP tuning}
\end{figure}

\begin{figure}[!htbp]
\centering
\includegraphics[width=.70\linewidth]{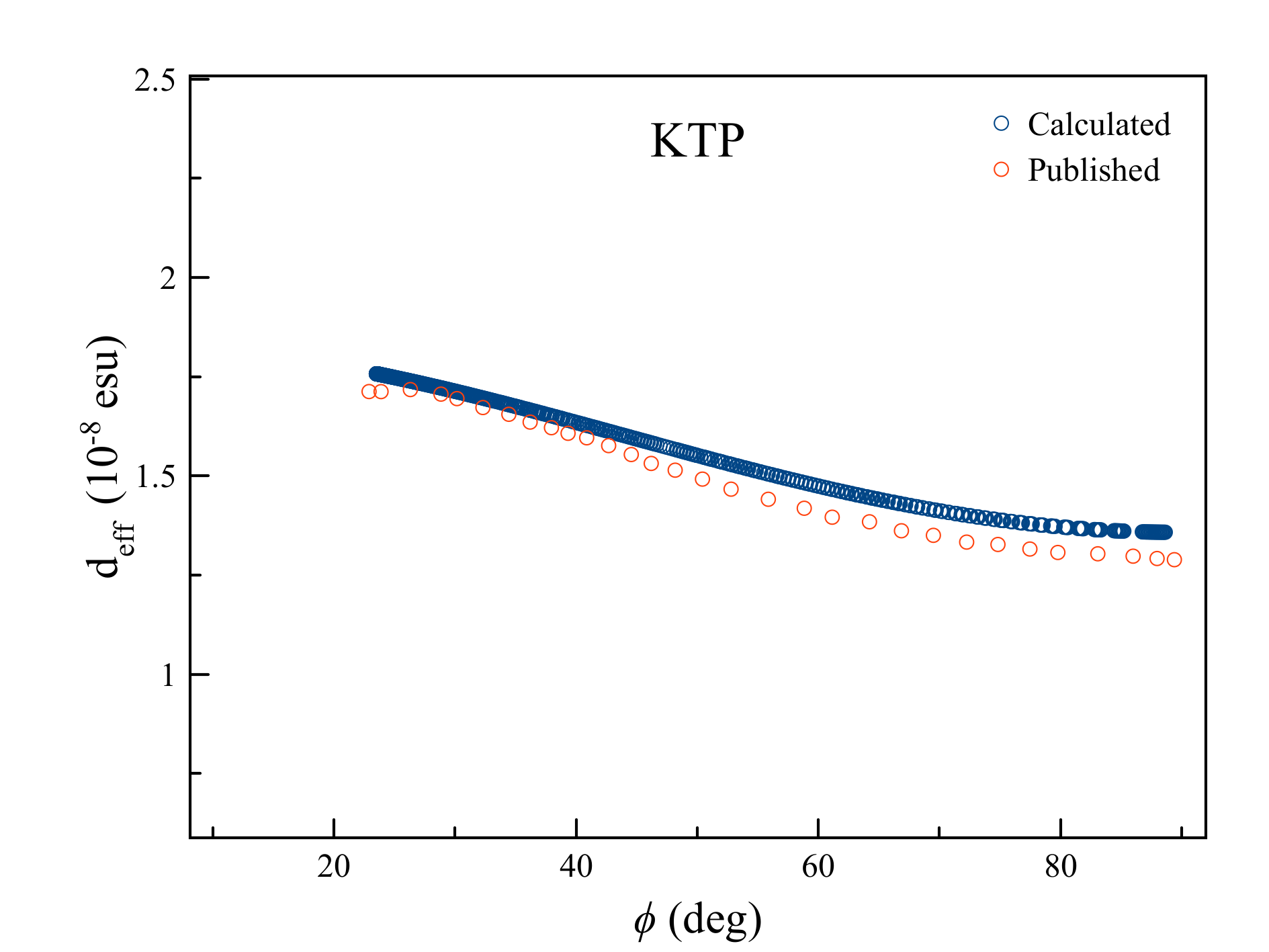}
\caption[KTP deff]{$d_{\rm eff}$ for KTP as a function of the azimuthal angle $\phi$, plotted in the first quadrant, where the blue dots are the calculated $d_{eff}$ curve and the red dots are the published values \cite{yao1984} for comparison.}
\label{fig:KTP deff}
\end{figure}

\bibliography{typeII_mof}
